\documentclass[twocolumn,preprintnumbers,amsmath,amssymb,superscriptaddress]{revtex4}

\usepackage{epsf}
\usepackage{epsf}
\usepackage{graphicx}
\usepackage{sidecap}
\usepackage{array}
\usepackage{amsmath}
\usepackage{amssymb}
\usepackage{dcolumn}
\usepackage{epstopdf}
\usepackage{bm}
\usepackage{amsmath}

\usepackage{color}

\def \beq {\begin{equation}}
\def \eeq {\end{equation}}
\begin{document}
	
\title {Observation of multiple Dirac states in a magnetic topological material EuMg$_2$Bi$_2$}
\author{Firoza~Kabir}
\affiliation {Department of Physics, University of Central Florida, Orlando, Florida 32816, USA}
\author{M.~Mofazzel~Hosen}
\affiliation {Department of Physics, University of Central Florida, Orlando, Florida 32816, USA}
\author{Fairoja~Cheenicode-Kabeer}
\affiliation{Department of Physics and Astronomy, Uppsala University, P.\ O.\ Box 516, S-75120 Uppsala, Sweden}
\author{Alex Aperis}
\affiliation {Department of Physics and Astronomy, Uppsala University, P.\ O.\ Box 516, S-75120 Uppsala, Sweden}
\author {Xiaxin Ding}
\affiliation{Idaho National Laboratory, Idaho Falls, ID 83402, USA}
\author{Gyanendra~Dhakal}
\affiliation {Department of Physics, University of Central Florida, Orlando, Florida 32816, USA}
\author{Klauss~Dimitri}
\affiliation {Department of Physics, University of Central Florida, Orlando, Florida 32816, USA}
\author{Christopher Sims}
\affiliation {Department of Physics, University of Central Florida, Orlando, Florida 32816, USA}
\author{Sabin Regmi}
\affiliation {Department of Physics, University of Central Florida, Orlando, Florida 32816, USA}
\author{Luis Persaud}
\affiliation {Department of Physics, University of Central Florida, Orlando, Florida 32816, USA}
\author{Krzysztof Gofryk}
\affiliation {Idaho National Laboratory, Idaho Falls, ID 83402, USA}
\author{Peter M. Oppeneer}
\affiliation {Department of Physics and Astronomy, Uppsala University, P.\ O.\ Box 516, S-75120 Uppsala, Sweden}
\author{Dariusz Kaczorowski}
\affiliation {Institute of Low Temperature and Structure Research, Polish Academy of Sciences, 50-950 Wroclaw, Poland}

\author{Madhab~Neupane}
\affiliation {Department of Physics, University of Central Florida, Orlando, Florida 32816, USA}

\date{\today}

\begin{abstract}
\noindent
	{Initiated by the discovery of topological insulators, topologically non-trivial materials, more specifically topological semimetals and metals have emerged as new frontiers in the field of quantum materials. In this work, we perform a systematic measurement of EuMg$_2$Bi$_2$, a compound with antiferromagnetic transition temperature at 6.7 K, observed via electrical resistivity, magnetization and specific heat capacity measurements. By utilizing angle-resolved photoemission spectroscopy in concurrence with first-principles calculations, we observe Dirac cones at the corner and the zone center of the Brillouin zone. From our experimental data, multiple Dirac states at $\Gamma$ and K points are observed, where the Dirac nodes are located at different energy positions from the Fermi level. Our experimental investigations of detailed electronic structure as well as transport measurements of EuMg$_2$Bi$_2$ suggest that it could potentially provide a platform to study the interplay between topology and magnetism.}
	
\end{abstract}
\maketitle
\noindent
\noindent
\textbf{INTRODUCTION}\\
The interplay between magnetic order and nontrivial topology can generate alluring topological quantum phenomena in magnetic topological compounds, such as the quantum anomalous Hall effect \cite{Yu1, K28, Lo}, axion insulator states \cite{T, M, N, Mogi, Zn, YF, CV, Sabin}, chiral Majorana fermions \cite{L} etc. Recently, enormous attempts have been made to embed magnetism into topological materials, such as realization of the Quantum Anomalous Hall (QAH) effect in magnetically-doped (Bi,Sb)$_2$Te$_3$ thin films \cite {Yu1, K28}. Though these types of doped magnetic materials are considered as good platforms to observe non trivial topology but usually the magnetic impurities initiate sharp inhomogeneity, which is considered to be one of the main reasons that the QAH usually develops at remarkably low temperatures (less than 100 mK), preventing further inspection of topological quantum effects \cite{K28}. An explicit solution to prevent this type of inhomogeneity is to explore intrinsic magnetic topological insulators (TIs) with magnetic order in the stoichiometric compositions. In the past years, some magnetic topological states have been theoretically introduced, such as dynamical axion field \cite{K40}, antiferromagnetic topological insulator \cite{Mong},  magnetic Dirac semimetal \cite{K32}, and magnetic Weyl semimetals \cite{K30}. Recently, intrinsic anomalous Hall conductivity (above room temperature) has been detected experimentally in the ferromagnetic kagome metal Fe$_3$Sn$_2$ \cite{Linda}. Additionally, large intrinsic anomalous Hall effect in the half-metallic ferromagnet Co$_3$Sn$_2$S$_2$ has also been discovered \cite{Qi}, where a pronounced peak at the Fermi level has been identified to appear from the kinetically frustrated kagome flat band by using scanning tunnelling microscopy \cite{J.Xin}. Recently, magnetic Weyl semimetal phase in ferromagnetic crystal Co$_3$Sn$_2$S$_2$ has been experimentally observed which may serve as a platform for realizing phenomena such as chiral magnetic effects, unusual large anomalous Hall effect and quantum anomalous Hall effect \cite{Y.chen}. More interestingly, the antiferromagnetic topological insulator (AFMTI) state has been realized in MnBi$_2$Te$_4$ \cite{Mam, Mong, S1, S2}. Recently, Dirac surface states in intrinsic magnetic topological insulator EuSn$_2$As$_2$ has been observed which provides clear evidence for nontrivial topology \cite{EuSn}. It would be very interesting to have a topological material possessing multiple Dirac states in order to study the interplay between magnetism and multiple Dirac fermions. 

For a long time, rare-earth (RE) intermetallic materials have achieved considerable interest due to enthralling properties at low temperatures including complex magnetic phases, valence fluctuations, heavy-fermion states and Kondo behavior \cite{K1,K3,K4,K6}. In particular, Eu based compound, EuMnBi$_2$ exhibits linear band-dispersion stemming from Dirac cones near the Fermi surface \cite{May}. Electronic structure calculations of EuCd$_2$As$_2$ revealed similar Dirac dispersion found in the nonmagnetic Dirac semimetal Cd$_3$As$_2$ \cite{Rahn, Neupane}. Recently, single pair of Weyl Fermions has been predicted in half-metallic semimetal EuCd$_2$As$_2$ \cite{EuCd2As2}. All those tremendous interesting  phenomena of the above mentioned Eu-based compounds have prompted us to investigate the detailed electronic structure of EuMg$_2$Bi$_2$ in order to elucidate possible interplay between magnetism and topology in this novel system.
\begin{figure*}
	\includegraphics[width=18 cm]{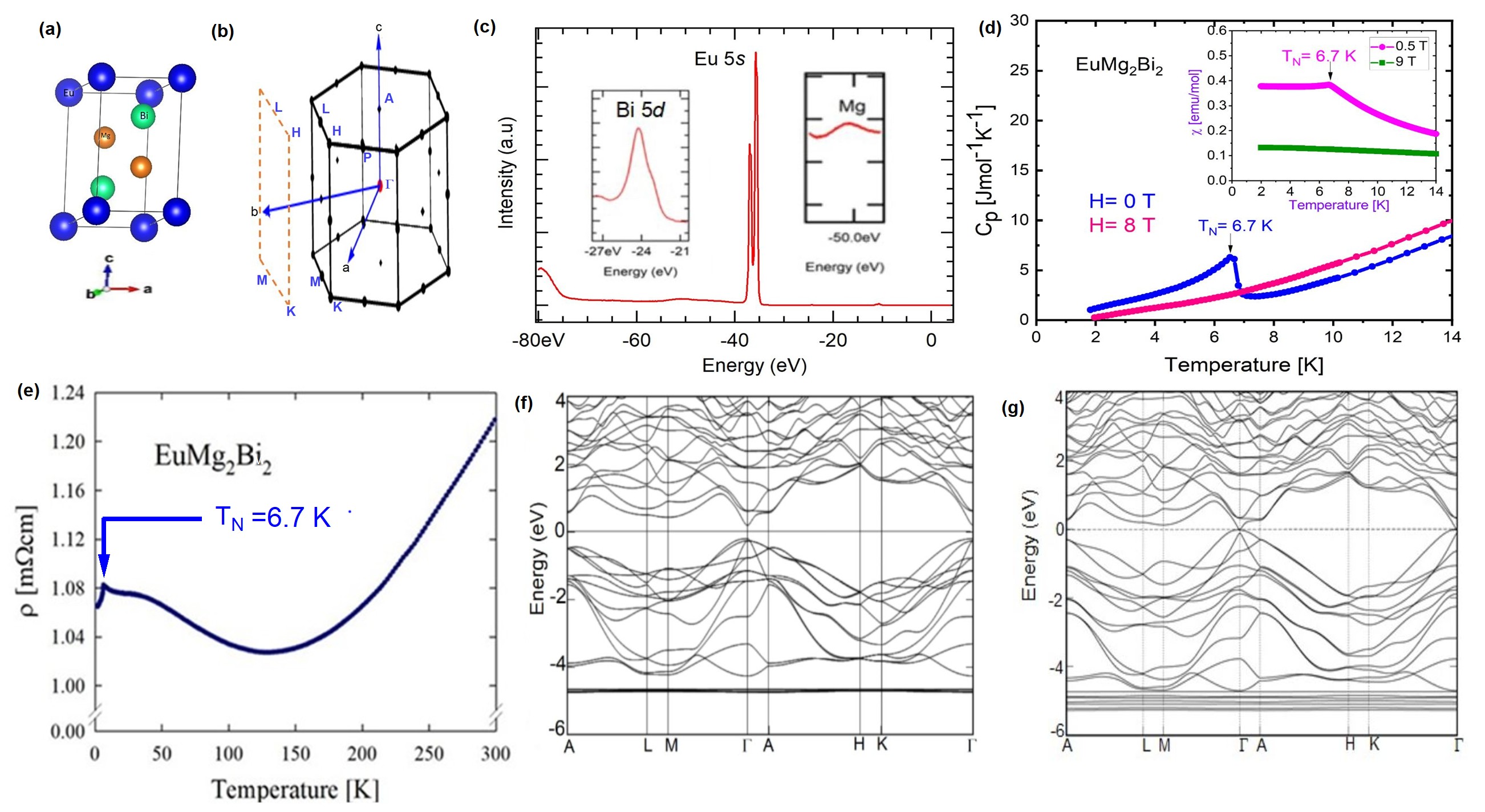}
	\caption{{\textbf{Crystal structure, sample characterization and electronic structure calculation of EuMg$_2$Bi$_2$}}. (a) Trigonal Crystal structure of EuMg$_2$Bi$_2$, where blue spheres represent Eu atoms, orange spheres represent Mg atoms and green spheres represent Bi atoms, as indicated in the figure. (b) Three dimensional (3D) bulk Brillouin zone of the crystal with its projection on the rhombohedral (010) surface. High symmetry points are marked on the plot. (c) Spectroscopic measured core levels of EuMg$_2$Bi$_2$; sharp peaks of Eu 5\textit{s}, Bi 5\textit{d} and Mg 2\textit{p} levels are observed which indicate good sample quality of our crystal. (d) Heat capacity of EuMg$_2$Bi$_2$ crystals measured in zero magnetic field (blue curve) and in high field of 9 T (pink curve). A pronounced anomaly marks an antiferromagnetic phase transition at 6.7 K. The inset shows the temperature dependence of the magnetic susceptibility of EuMg$_2$Bi$_2$  with antiferromagnetic transition at \textit{T$_N$} = 6.7 K. Measurements were performed in a field of 0.5 T (violet curve) and 9 T (green curve). (e) Temperature variation of the electrical resistivity of EuMg$_2$Bi$_2$, measured within the trigonal plane. The arrow marks the temperature of antiferromagnetic phase transition. (f) Density-functional theory $+U$ band structure calculations of EuMg$_2$Bi$_2$ (without spin-orbit coupling (SOC)) along various high-symmetry directions (AFM state). (g)  Calculated band structure of EuMg$_2$Bi$_2$ with SOC (AFM state). High symmetry points are also marked.}
\end{figure*}


  \begin{figure*}
  	\centering
  	\includegraphics[width=20 cm]{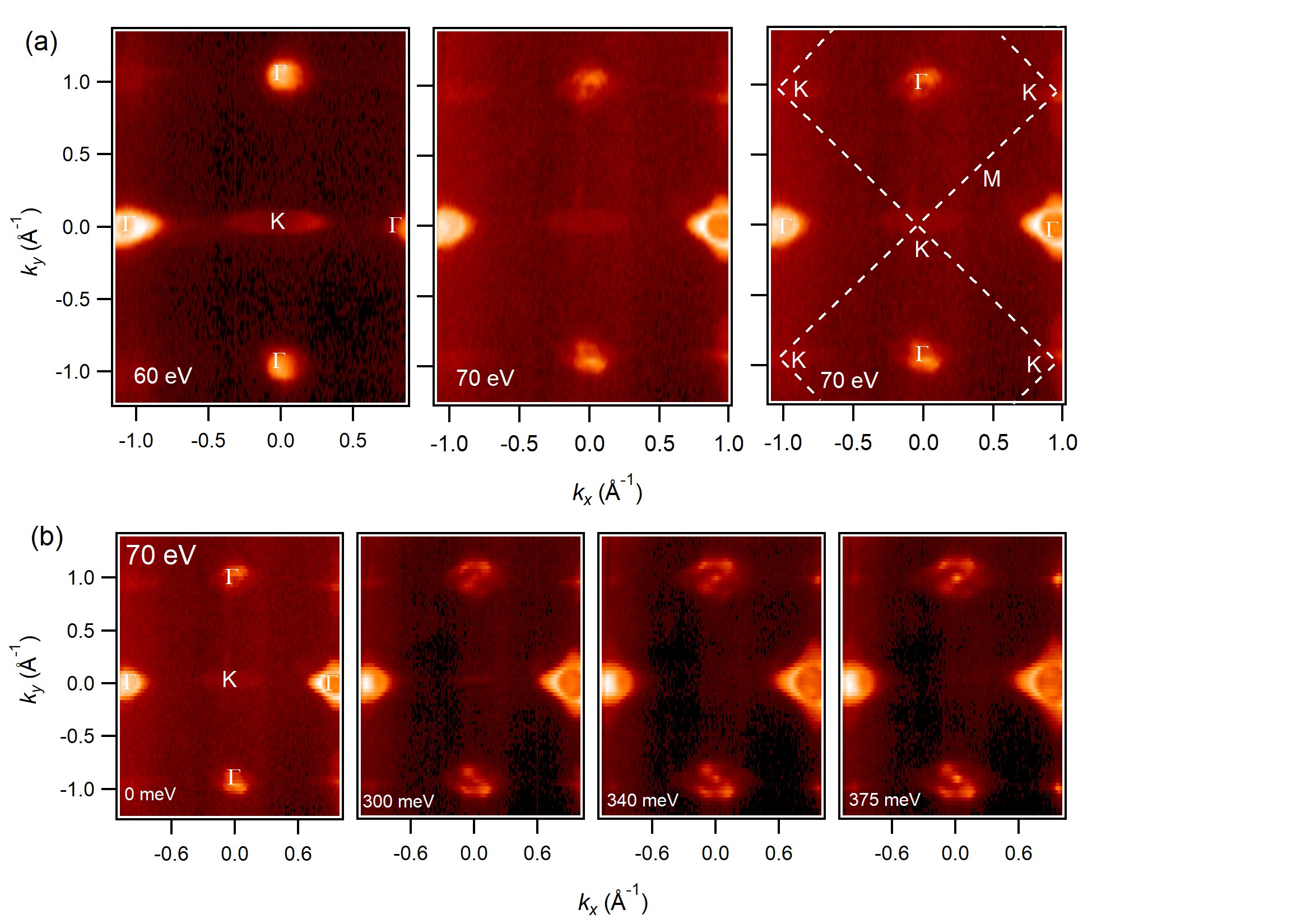}
  	\caption{{\textbf{Fermi surface map and constant energy contours of EuMg$_2$Bi$_2$}.} (a) Experimentally measured Fermi surface maps at photon energy of 60 and 70 eV with indication of various high symmetry points of the Brillouin zone (right most panel). (b) Constant energy contour plots at various binding energies for the photon energy of 70 eV. Binding energies are noted in the plots. All measurements were performed at the ALS beamline 10.0.1 at a temperature of 15\,K. }
  \end{figure*}
  Hence, to understand the interplay between topology and magnetism in the Eu-based 122 family, we present a systematic study of the intrinsic magnetic topological material EuMg$_2$Bi$_2$, using ARPES, transport, magnetization and thermodynamic properties measurements together with first-principles calculations. Here, we report a comprehensive investigation of the bulk physical properties and electronic structure of EuMg$_2$Bi$_2$. Our thermodynamic (magnetization, heat capacity) and transport (electrical resistivity) measurements have revealed an anti-ferromagnetic phase transition at 6.7 K in this compound. By means of angle resolved photoemission spectroscopy (ARPES), we have measured the electronic structure of this system. The spectroscopic studies have been supplemented by first-principles band structure calculations, the main finding is the discovery of multiple fermionic surface states with numerous Dirac crossings in this compound. While multiple Dirac states have recently been observed by ARPES in a couple of topological compounds, like non-magnetic strong topological metals, Zr$_2$Te$_2$P \cite{cava} and Hf$_2$Te$_2$P \cite{Dolar}, crystallizing with  a tetradymite structure, up to date, no Eu-based 122 pnictide nor any other magnetic 122 compound with trigonal crystal structure has experimentally been reported to exhibit multiple Dirac states. Therefore, our experimental observation of the multiple Dirac features in the antiferromagnetic bismuthide EuMg$_2$Bi$_2$ opens a new door for the exploration of new quantum phases in magnetic materials, which can result in understanding the tempting issue of the interplay of magnetism and topology. \\~\\
 \textbf{RESULTS}\\
 \textbf{Crystal structure and sample characterization} \\
 EuMg$_2$Bi$_2$ crystallizes in the CaAl$_2$Si$_2$ structure type (trigonal, No.\ 164, P3m1), as shown in Figure.\ 1(a). This structure is quite dissimilar than the common body-centered tetragonal ThCr$_2$Si$_2$-type structure of 122 compounds \cite{33}, for an example, the EuMg$_2$Bi$_2$ crystal possesses the largest geometric anisotropy and strongest hybridizations, which have been detected between Bi and Eu, explaining the small band gap and large hole mobility in this compound \cite{30}. The lattice parameters of EuMg$_2$Bi$_2$ have been previously reported as \textit{a} = 4.7771(1) {\AA} and \textit{c} = 7.8524(1) {\AA} \cite{May}. The 3D bulk Brillouin zone of the crystal is visualized in Figure.\ 1(b), where the projection on the rhombohedral surface is shown by brown lines. Figure.\ 1(c) presents the core level photoemission spectrum of EuMg$_2$Bi$_2$ crystal, which clearly manifests the characteristic peaks coming from Eu \textit{5s}, Bi \textit{5d}, and Mg \textit{2p} orbitals.

 \begin{figure*}
 \centering
 \includegraphics[width=17.0cm]{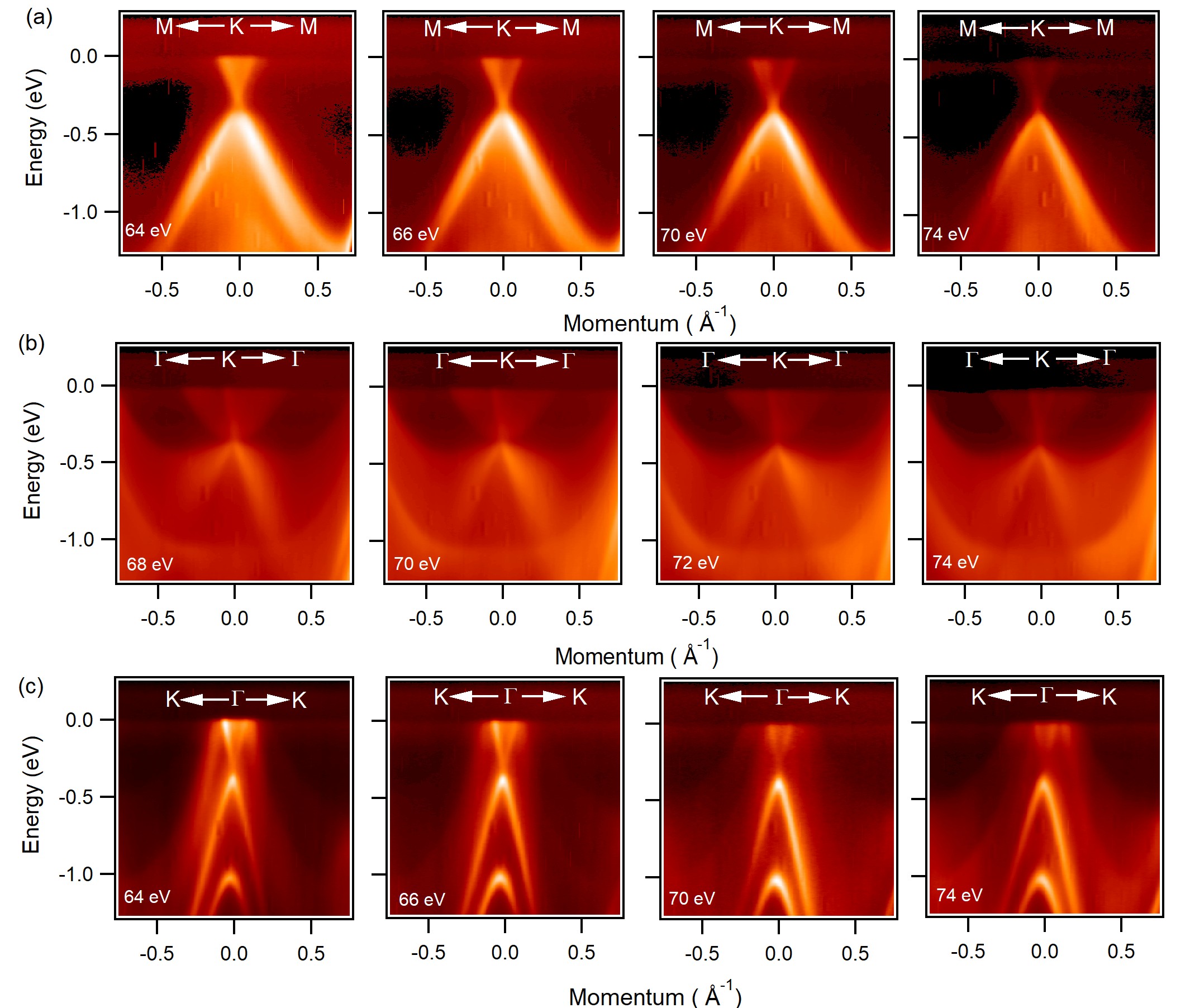}
 \caption{{\textbf{Observation of multiple Dirac states in EuMg$_2$Bi$_2$}.} (a) ARPES measured dispersion maps along the M-K-M direction using various photon energies. Linearly dispersive states are observed that do not show dispersion with a dependence on photon energy and Dirac cone is observed at the K point. (b) Dispersion maps along the high symmetry $\Gamma$-K-$\Gamma$ direction. (c) Photon energy dependent band dispersion along the K-$\Gamma$-K direction. Here, a Dirac cone at the $\Gamma$ point is observed experimentally. Measured photon energies are noted in the plots. All the measurements were performed at the ALS end station 10.0.1 at a temperature of 15 K.}
 \end{figure*}
 Figure 1(d) shows the temperature dependence of the heat capacity of single crystalline EuMg$_2$Bi$_2$  measured in zero and 8 T magnetic fields. A sharp $\lambda$- like peak is observed at the antiferromagnetic phase transition, \textit{T}$_N$ = 6.7 K, in agreement with the previous results \cite{May,Andrew}. Applying a magnetic field stronger than 3 T suppresses the magnetic ordering in this material (see also Ref.\ \cite{May, Andrew}). The inset presents the low temperature magnetic susceptibility of EuMg$_2$Bi$_2$ measured in magnetic fields of 0.5 T and 9 T. A characteristic anomaly (as expected for antiferromagnetic systems) is observed at the Neel temperature. Both measurements confirm the existence of a bulk antiferromagnetic ordering below 6.7 K in the EuMg$_2$Bi$_2$ crystals. Figure. 1(e) shows the temperature dependence of the electrical resistivity of EuMg$_2$Bi$_2$ measured with electric current flowing within the trigonal plane. At low temperatures, a sharp anomaly at 6.7 K marks the antiferromagnetic transition, in accord with thermodynamic measurements. Increasing the temperature above \textit{T}$_N$, the resistivity starts to decrease and forms a shallow minimum at 125 K. The magnitude and overall temperature dependence of the electrical resistivity in the paramagnetic state are typical for the characteristics of magnetic semimetals or narrow gap semiconductors.\\
 Figures 1(f) and 1(g) show the band structures of EuMg$_2$Bi$_2$ along the high symmetry directions of the Brillouin zone for the anti ferromagnetic phase calculated without and with spin-orbit coupling (SOC), respectively. From first-principles calculations of  EuMg$_2$Bi$_2$, the band gap (without SOC) between the valence and conduction bands has been previously reported as 0.25 eV \cite{30}.  Here, we employ the Density Functional Theory (DFT) $+U$ approach, which provides better localized $4f$ states than the common DFT approach.  With the DFT$+U$ approach, the occupied Eu $4f$ states are quite localized at a large binding energy of $\sim$5 eV. Strong hybridizations between Bi and Eu atoms explain the small band gap ($\sim$ 0.45 eV) and large hole mobility in EuMg$_2$Bi$_2$ \cite{30}. 
 \\~\\
\textbf{Fermi surface and constant energy contour plots of EuMg$_2$Bi$_2$}\\
In order to visualize the evolution of band structures with the binding energies, we discuss the experimental electronic structures of EuMg$_2$Bi$_2$ by studying the Fermi surface and constant energy contours. For this purpose, we unveil the detailed electronic structure of  EuMg$_2$Bi$_2$ by ARPES measurements, which are shown in Figures.\ 2 and 3. Figure. 2(a) presents the broad mapping of the Fermi surfaces of EuMg$_2$Bi$_2$  covering multiple Brilliouin zones (BZs), where the periodicity in the mapping can be clearly seen. The Fermi surface mappings at various photon energies (60 eV and 70 eV) are presented, where the Fermi pockets are indicated according to the high symmetric point of the Brillouin zone (BZ). Right most panel of Figure.\ 2(a) shows the Fermi surface mapping measured at photon energy 70 eV, where the dashed white squares indicate the BZ of EuMg$_2$Bi$_2$ system. Constant energy contours at various binding energies are displayed in Figure.\ 2(c). In the Fermi surface of EuMg$_2$Bi$_2$, we observe a small diamond shaped Fermi pocket  with a central ring shape pocket around the $\Gamma$ point and two intersecting elliptical shaped Fermi pockets at the K point of the BZ (photon energy 70 eV). Moving toward higher binding energies, the diamond shaped Fermi pocket at the $\Gamma$ point expands and the elliptical shaped pocket at the K point starts to decrease in size. The $\Gamma$ point is surrounded by linearly dispersive band as well as hole band. As we move further up to higher binding energy of 375 meV, we observe the Dirac node at the $\Gamma$ point and we notice Dirac node at the K point located 340 meV below the Fermi level. With increasing binding energy, the Dirac state at the K point starts to shrink at a point but we can see some large diamond shaped features with inner circle at the $\Gamma$ point. Hence, from this overall Fermi surface mapping, we can see the multiple Fermi pockets at the $\Gamma$ and K point. To confirm these Fermi pockets, we perform 2$^{nd}$ derivative analysis of  Fermi surface and constant energy contours presented in supplementary figure 1.\\~\\
\textbf{Observation of multiple Dirac states in EuMg$_2$Bi$_2$}\\ 
  To reveal the origin of states around the $\Gamma$ and K points, we perform photon energy dependent ARPES measurements as shown in Fig.\ 3. The linearly dispersive states which form the Dirac like states in our measured ARPES spectra, do not disperse with the photon energy, suggesting that they are surface originated (see figure 3). And we do not see any irrelevant band along M-K-M high symmetry direction. Additionally, Figs.\ 3(b) and 3(c) show the dispersion maps along the $\Gamma$-K-$\Gamma$ and K-$\Gamma$-K direction at various photon energies ranging from 64 to 74 eV. These photoemission spectra reveal that the Dirac cone at $\Gamma$ do not disperse with photon energies. 
  We can see the Dirac point at 340 meV for the K point and at around 375 meV for the $\Gamma$ point. Second derivative plots are presented in supplementary figures 2, 3 and 4 to confirm the multiple Dirac states more clearly. Therefore, our experimental data showing gapless linearly dispersive  surface states in the vicinity of the $\Gamma$ and K points confirm the presence of the multiple Dirac-like states in EuMg$_2$Bi$_2$. \\~\\
We perform systematic ARPES measurements of EuMg$_2$Bi$_2$ crystals, covering a large area of the Brillouin zone. Our measurements reveal the presence of multiple Fermi pockets in the Brillouin zone enclosing the high symmetry points. Specifically, Dirac dispersions are observed around the $\Gamma$ and K points of the surface Brillouin zone, where the energy of the Dirac points are located at 375 and 340 meV below the Fermi level, respectively. Dirac points can be accessed by the chemical doping and electrical gating in this system. Dirac dispersion at the K point (along M-K-M direction) is observed to be well separated from other irrelevant bands in the vicinity of the Fermi level in EuMg$_2$Bi$_2$. Although the \textit{f}-electron band is located well below the Fermi level ($\sim$ 4.5 eV), the presence of the multiple Dirac states near the Fermi level makes this system interesting to study the interplay between Dirac fermion and magnetism. Altogether, our detailed studies of the electronic structure, transport and thermodynamic measurements of intrinsic magnetic topological compound EuMg$_2$Bi$_2$ provide an ideal platform to understand the interplay between magnetism and non-trivial topology and thus one can gain deeper interpretation of magnetic topological materials from this system.\\~\\
 {
\textbf{METHODS}\\
\textbf{Sample growth and characterizations}\\
 Single crystals of EuMg$_2$Bi$_2$ were grown by the Sn flux technique as described elsewhere \cite{43}. The crystal structure was deduced from X-ray diffraction on a Kuma-
 Diffraction KM4 four-circle diffractometer equipped with a CCD camera using Mo K$\alpha$ radiation. Chemical composition of the crystals were inspected by energy dispersive X-ray analysis by using a FEI scanning electron microscope equipped with an EDAX Genesis XM4 spectrometer. The electrical resistivity, heat capacity, and magnetic susceptibility were measured using a Quantum Design PPMS system, equipped with 9 T superconducting magnet and using ACT, HC, and VSM options respectively.
 \\~\\
 \textbf {Synchrotron measurements}\\
  Synchrotron-based ARPES measurements of the electronic structure of EuMg$_2$Bi$_2$ were performed at ALS BL 10.0.1 with a Scienta R4000 hemispherical electron analyzer. The samples were cleaved in situ under ultra high vacuum conditions (5x10\textsuperscript{-11} Torr) at 15 K. The energy resolution was set to be better than 20 meV and the angular resolution was set to be finer than  0.2$^{\circ}$ for the synchrotron measurements.The EuMg$_2$Bi$_2$ specimens were found to be very stable and did not display any signs of derogation for the typical measurement period of 20 hours. The crystals were cut into flat small pieces, mounted on copper posts and on the top of the sample, ceramic posts were also attached by using silver epoxy, and then the samples were loaded into the measurement chamber.\\~\\
 \textbf{Electronic structure calculations}\\
The electronic structure calculations and structural optimization were carried out within the density-functional formalism as implemented in the Vienna \textit{ab initio} simulation package (VASP) \cite{Kresse_1, Kresse_2}.
Exchange and correlation were treated within the generalized gradient approximation (GGA) using the parametrization of Perdew, Burke, and Ernzerhof (PBE) \cite{GGA_1} with and without spin-orbit interaction. The projector-augmented wave (PAW) method \cite{PAW, Blochl} was employed for the wave functions and pseudopotentials to describe the interaction between the ion cores and valence electrons.The lattice constants and atomic geometries were fully optimized and obtained by minimization of the total energy of the bulk system. Two primitive cells repeated along the $z$ axis gave the lowest total energy and were used as AFM configuration. We used a kinetic energy cutoff of 520 eV, a $\Gamma$-centered Monkhorst-Pack \cite{Monkhorst}  (9x9x3) k-point mesh, and a Gaussian smearing of 0.05 eV. To account for the strongly localized Eu $4f$ orbitals, we used an onsite Coulomb parameter $U=11.0$ eV. The atomic basis positions and unit cell vectors are fully relaxed for the AFM configuration, until the magnitude of the force on each atom is reduced to below $10^{-5}$ eV/{\AA}.\\~\\


\textbf{ACKNOWLEDGEMENTS}\\
M.N.\ is supported by the Air Force Office of Scientific Research under Award No. FA9550-17-1-0415 and the Center for Thermal Energy Transport under Irradiation, an Energy Frontier Research Center funded by the U.S. DOE, Office of Basic Energy Sciences. K.G acknowledges support from the DOE’s Early Career Research Program. X.D acknowledges support from INL’s LDRD program (19P45-019FP). F.C.-K., A.A., and P.M.O.\ acknowledge support from the Swedish Research Council (VR), the K.\ and A.\ Wallenberg Foundation (Grant No.\ 2015.0060) and the Swedish National Infrastructure for Computing (SNIC). We thank Sung-Kwan Mo for beamline assistance at the LBNL.\\~\\


\def\bibsection{

\end{document}